\documentstyle[12pt]{article}
\def\ssubset{\hspace{0.2em}{\subset
\kern-0.83em\raise0.27ex\hbox{${\scriptscriptstyle \subset}$}}\hspace{0.3em}}
\def\RA{${\bf R}^n$ }
\def\CA{${\bf C}^n$ }

\def\sab{$S_\alpha^\beta$ }
\def\sabu{$S_\alpha^{\beta}(U)$ }

\def\sao{$S_\alpha^0$ }
\def\sabk{$S_\alpha^{\beta}(K)$ }
\def\saok{$S_\alpha^{0}(K)$ }
\def\saou{$S_\alpha^{0}(U)$ }

\def\TMP{Theor.\ Math.\ Phys. }
\def\LMP{Lett.\ Math.\ Phys. }
\def\JMP{J.\ Math.\ Phys. }
\def\CMP{Commun.\ Math.\ Phys. }
\def\SM{Soloviev,~M.~A.: }
\def\plu{plurisubharmonic }

\begin{document}

\thispagestyle{empty}

\baselineskip=3ex

\noindent P.~N.Lebedev Institute Preprint \hfill FIAN/TD/24 - 95\\
 \noindent I.E.Tamm Theory Department  \hfill hep-th/9601005

\vspace{2cm}

\begin{center}
{\Large\bf
 An Extension of Distribution Theory Related to Gauge
Field Theory}

\bigskip

{\large M.~A.~Soloviev\footnote{e-mail: soloviev@lpi.ac.ru}

\bigskip

{\it Department of Theoretical Physics,
P.~N.~Lebedev Physical Institute, Leninsky prosp. 53, Moscow
117924, Russia}}

\end{center}

\vspace{3cm}


{\bf Abstract}  We show that a considerable part of the theory
of (ultra)distributions and hyperfunctions can be
extended to more singular generalized functions, starting from
an angular localizability notion introduced  previously.
 Such an extension is needed to treat gauge quantum field
theories with indefinite metric in a generic covariant gauge.
Prime attention is paid to the generalized
functions defined on the Gelfand-Shilov spaces \sao
which gives the widest framework for construction of gauge-like
models. We associate a similar test function space with
every open and every closed cone, show that these
spaces are nuclear and obtain the required formulas for their
 tensor products.  The main results include the generalization
of the Paley--Wiener--Schwartz theorem to the case of arbitrary
singularity and the derivation of the relevant  theorem on
holomorphic approximation.

\newpage

\setcounter{page}{1}

\section{Introduction}

This article will present a systematic development of the
generalized distribution-theoretic formalism proposed
in~\cite{S4} for the treatment of gauge quantum field
theories with indefinite metric. Our research was
motivated by the works [32, 21--23], where
several explicitly soluble field models have been analyzed,
whose correlation functions, when treated in a generic
covariant gauge, exhibit so singular infrared behavior that they
are neither tempered distributions or even hyperfunctions in
momentum representation. The suitable momentum-space test
functions proved to be in the Gelfand--Shilov~\cite{GS}
spaces \sab , where $\alpha$  and $\beta$ are model-dependent
and the superscript is less than one. These spaces have long ago
been exploited~\cite{E2,FS1}, equally with the spaces $S^\beta$
[2--5], in nonlocal QFT, where they came
into play in configuration representation since one was
concerned with ultraviolet singularities. The elements of \sab
decrease at infinity exponentially with order $1/\alpha$ and a
finite type, and their Fourier transforms behave analogously but
with order $1/\beta$. When $\beta<1$, these elements are entire
analytic functions and hence the notion of support loses its
sense for the general disrtibutions\footnote{Throughout this
paper, the term distribution is used synonymously with, and
instead of, generalized function, while the usual distributions
with finite order of singularity are called Schwartz or
tempered.} defined on \sab.  For this reason, formulating such
physical requirements as the spectral condition or (in the
ultraviolet singular case) causality becomes problematic.
Nevertheless, in the nonlocal QFT [2--8] a way of handling so
singular distributions has been found. In our opinion, the main
point is that these retain the angular localizability property.
In more detail, a distribution on \sab may be thought of as
carried by a closed cone $K$ if it allows a continuous linear
extension to the space \sabk associated naturally with this
cone, and then the following theorem holds.

{\bf Theorem 1}. {\it Let $K_1$ and $K_2$ be closed cones in
\RA and let  $\beta<1$. Every distribution on \sab carried by
$K_1\cup K_2$ can be decomposed into a sum of two distributions
carried by $K_1$ and $K_2$. Furthermore, if each of cones $K_1$,
$K_2$ is a carrier of $f\in S^{\prime \beta}_{\alpha}$, then so
is  $K_1\cap K_2$ and consequently $f$ has a unique minimal
carrier cone}.

For $0<\beta<1$, these facts can be established in an
elementary way~\cite{S3} which, however, is
inapplicable to the case $\beta=0$ when the Fourier transforms
of test functions have compact supports.  This borderline case
is of particular interest because using \sao  imposes no
restrictions on the singularity.  In other words, these spaces
may serve as a universal object for covariant formulation of
gauge QFT and construction of gauge-like models.  The general
proof of Theorem 1 set forth in~\cite{S4} and covering $\beta=0$
is based on a representation of the spaces \sabk as the inductive
limits of Hilbert spaces of analytic functions, which enables
one to take advantage of H\"{o}rmander's
$L^2$--estimates~\cite{H1} for solutions of the nonhomogeneous
Cauchy-Riemann equations. So the situation somewhat copies that
of the familiar case $\beta \geq 1$ : The local properties of
ultradistributions ($\beta>1$) and hence those of the Jaffe
quantum fields~\cite{J1,J2} allow a simple description by the
usual "partition of unity" method while the treatment of
hyperfunctions ($\beta=1$) and the Nagamachi--Mugibayashi
fields~\cite{BN,NM1,NM2} requires making recourse to the same
estimates or alternative methods of complex analysis~\cite{H2}.
Here we use the language of complex variables from the very
outset, and we refer to our previous work~\cite{S4} for its
relationship to the original Gelfand and Shilov definition of
\sab  used traditionally in nonlocal QFT and for the interplay
with the theory of hyperfunctions. First we assign a kindred
space to every open cone with vertex at the origin.  If
$U\subset {\bf R}^n$ is such a cone, then \sabu is defined to be
the union or, more precisely, the inductive limit of Hilbert
spaces $H^{\beta, b}_{\alpha, a}(U) \quad (a,b\rightarrow
\infty)$ consisting of entire functions on \CA and provided with
the scalar products

$$
{\langle \varphi,\psi \rangle}_{U,a,b} = \int
\bar{\varphi}(z)\psi (z)\exp\{-2\rho_{{}_{U,a,b}}(z) \}{\rm
d}\lambda, \eqno{(1)}
$$

\noindent
where ${\rm d}\lambda$ stands for the Lebesgue measure on \CA
and

$$
\rho_{{}_{U,a,b}}(z)\>=\> -\vert x/a\vert^{1/\alpha} +
d(bx,U)^{1/(1-\beta)} + \vert by\vert^{1/(1-\beta)} ,\eqno{(2)}
$$

\noindent
with $z=x+iy$  and $d(\cdot , U)$ being the distance of the point
to $U$. We remark that the union is independent of the choice of
the norm $|\cdot|$ in \RA and this degree of freedom will be of
use in what follows, though at first one may hold it
Euclidean.  A sequence $\varphi_\nu \in$\sabu is regarded to be
convergent if it is contained in some $H^{\beta, b}_{\alpha,
a}(U)$ and is $\|\cdot \|_{U,a,b}$--convergent. Starting from
\sabu, the space \sabk associated with a closed cone $K$ is
constructed by taking another inductive limit through
those $U$ which contain the set $K\setminus \{0\}$   and shrink
to it.  The origin plays the role of the least element in
the lattice of closed cones, and their associated space
\sab(\{0\}) is defined by the same formulas, but with the first
term in (2) being dropped and $|\cdot|$ substituted for $d(\cdot
, U)$. It should be noted that the designations used in (2) are
inherited from nonlocal QFT and will be convenient for the most
derivations below.  However, in  the final application to
infrared singular fields, the replacement $x\rightarrow p, \quad
y\rightarrow q$ is advisable of course.

For reader's convenience, in Sec. II we briefly sketch the proof
of basic Theorem 1 presented in every detail in~\cite{S4}.
We show also that the spaces associated with cones are nuclear.
This extends, with a simpler proof, Mitiagin's  well-known
 result concerning \sab and implies some nice
topological
 properties which are of use in QFT. Specifically,
the nuclearity enables us to prove in Sec. III that the tensor
product of the spaces over open cones, when being completed
under a proper topology, is identical to the space associated
with the cross product of the cones. Combining this fact with
Theorem 1 and Theorem 11.5 of~\cite{K2}, we obtain in
Sec.  IV an extension of the Paley--Wiener--Schwartz
theorem which relates the support properties of
distributions to the growth properties of their Fourier--Laplace
transforms and is of prime importance in constructive
QFT. All these derivations are expounded for the
widest distribution class $S^{\prime 0}_\alpha$ when,
fortunately, the designations get simplified.  Certainly the
case of nonzero $\beta<1$ can be treated in the same manner but
the corresponding theorem of PWS type has already been proved
in~\cite{S3} by another method.  Sec. V contains a complete
proof of an approximation theorem announced in~\cite{S4}, which
asserts that the space \sao is dense in every \saok.  Here again
a leading part is played by H\"{o}rmander's $L^2$--estimate but
this time relating to the dual equation.  This theorem implies
 that the space of distributions carried by
$K$ can be identified with $S^{\prime 0}_\alpha (K)$. The listed
results form a basis for more special derivations such as
structure theorems and representations for the infrared
singular correlation functions and enable one to extend methods
of Euclidean and constructive field theory  to gauge
models with such a behavior. Sec.  VI is devoted to concluding
remarks.

\section{Nuclearity}

{\bf Theorem 2.} {\it For each open cone $U$, the space
 \saou is nuclear and this property is inherited by the spaces
corresponding to closed cones.}

{\it Proof.} We refer the reader to Schaefer~\cite{S}
for definition and basic facts concerning nuclear locally convex
spaces. It is sufficient to show that, for each $a^\prime>a,
\quad b^\prime>b$, the inclusion mapping
$i^{bb^\prime}_{aa^\prime}:\quad H^{0,b}_{\alpha,a}(U)\rightarrow
H^{0,b^\prime}_{\alpha,a^\prime}(U) $ is a Hilbert--Schmidt
operator. Then $i^{bb^\prime}_{aa^\prime}$ is also nuclear since
it may be regarded as a composition of two Hilbert--Schmidt
operators. According to~\cite{S}, it follows that the
projective limit

$$
S_{\alpha,a+} ^{0,b+}(U) = \bigcap_{\varepsilon >0}
H^{0,b+\varepsilon}_{\alpha,a+\varepsilon}(U)
\eqno{(3)} $$

\noindent
is a nuclear space. The spaces \saou and \saok can be
represented as countable inductive limits of auxiliary spaces of
the form (3) and so they are nuclear by the inheritance
properties listed in~\cite{S}, Sec. III.7.4.

We denote by ${\cal H}^{0,b}_{\alpha,a}(U)$ the Hilbert space of
locally square-integrable functions on \CA equipped with the
same scalar product as that of $H^{0,b}_{\alpha,a}(U)$, i.e.,
${\cal H}^{0,b}_{\alpha,a}(U)= L^2({\bf C}^n, e^{-2\rho}{\rm
d}\lambda)$. Let us consider within this scale of spaces the
integral operator $G$ defined by the kernel

$$
  G(z^\prime-z) = {1\over (2\pi)^{2n}}\int
  \frac{e^{ip(x^\prime-x) + iq(y^\prime-y)}}{(p^2+ q^2
  +M^2)^{n+1}}{\rm d}^n p\,{\rm d}^n q, \eqno{(4)} $$

\noindent
which is nothing but the inverse of $(-\Delta
+M^2)^{n+1}$.  We claim that, if $M$ is properly chosen, $G$
acts as a Hilbert--Schmidt operator ${\cal
H}^{0,b}_{\alpha,a}(U)\rightarrow {\cal
H}^{0,b^\prime}_{\alpha,a^\prime}(U)$. On the other hand, its
restriction onto $H^{0,b}_{\alpha,a}(U)$ is the identity
operator, up to the factor $M^{-2(n+1)}$, because the analytic
functions satisfy the Laplace equation.
Thus showing our claim
implies that the injection $i^{bb^\prime}_{aa^\prime}$ has the
desired property. We use the standard estimate

$$
|G(z^\prime-z)| \leq C\exp\{-m|z^\prime -z|\}
\eqno{(5)}
$$ valid for any $m<M$. Due to the rotational invariance,
 when deriving (5) one
may assume  that the only nonzero component
of the argument is $x^\prime_1 -x_1$ and then
shift the path of integration in the
$p_1$--plane.  Changing variables and combining (5) with the
elementary inequalities $\sqrt{2}|z|\geq |x|+|y|$, and

$$
|y^\prime -y|\leq |y^\prime |+ |y|,\quad  d(x^\prime -x,U)\leq
d(x^\prime,U)+|x|, \quad -|x^\prime -x|^{1/\alpha}\leq
|x|^{1/\alpha} -|x^\prime |^{1/\alpha}, \eqno{(6)} $$ \noindent
where the latter is true since $\alpha >1$, we find that
$$
\int |G(z^\prime -z)|^2 \exp\{2\rho_{{}_{U,a,b}}(z)\}{\rm
d}\lambda \leq C^\prime \exp\{2\rho_{{}_{U,a,b}}(z^\prime)\}
\eqno{(7)}
$$
\noindent
provided $M>b\sqrt{2}$. Then the integral $\int G(z^\prime
-z)\varphi (z){\rm d}\lambda$, where $\varphi \in {\cal
H}^{0,b}_{\alpha,a}(U)$, belongs to any ${\cal
H}^{0,b^\prime}_{\alpha,a^\prime}(U)$ with $a^\prime>a, \quad
b^\prime>b$ by virtue of Schwarz's inequality. Next we note
that multiplication by the weight function $e^{-\rho}$ generates
a unitary mapping of ${\cal H}^{0,b}_{\alpha,a}(U)$ onto $L^2
({\bf C}^{n})$ and hence our claim is equivalent to saying
that $e^{-\rho^\prime}Ge^\rho$ is a Hilbert--Schmidt operator on
$L^2$. The same formula (7) shows that
this is the case  with $M$ as above, i.e., the kernel of $e^{-\rho^\prime}Ge^\rho$  is
square-integrable. For
accuracy, let us specify a space containing those in
question whereon $(-\Delta +M^2)^{n+1}$
acts as an automorphism. We take it to be the dual of the space
$S_{1,l}({\bf R}^{2n})$ consisting of infinitely differentiable
functions with the property that

$$
|\partial^\kappa \varphi(x,y)| \leq C_\kappa
\exp\{-|x/l|-|y/l|\}
$$

\noindent
for all multi-indices $\kappa$, and equipped with the
corresponding topology. Clearly $(-\Delta +M^2)^{n+1}$ is
a continuous operator on $S_{1,l}$, while $G$ maps this space
into itself provided $l>\sqrt{2}/M$ and then it is just the
inverse operator as can easily be seen with the use of Fubini's
theorem. The dual space contains ${\cal
H}^{0,b^\prime}_{\alpha,a^\prime}(U)$ if
$l<1/b^\prime$ and the dual operator also has a continuous
inverse whose restriction to ${\cal H}^{0,b}_{\alpha,a}(U)$
cannot be different from $G$ since $S_{1,l}$ is dense therein.
This completes the proof.

{\it Remark 1}. Each Hilbert--Schmidt operator is compact and
therefore so is the injection $i^{bb^\prime}_{aa^\prime}$. The
limit space of a projective (injective) sequence of locally
convex spaces with compact  connecting mappings is referred to
as an FS (DFS) space respectively, FS being abbreviation of
Fr\'echet--Schwartz and D signifying "dual". Thus, as a
consequence of Theorem 2, the spaces \saou and \saok are DFS
whereas their strong dual spaces as well as the spaces (3) are
FS. In this connection, it perhaps should be recalled that all
spaces of these two types are complete, reflexive, separable and
Montel, see, e. g.,~\cite{KO} for more detailed comments.

  Certainly, these nice
topological properties hold true~\cite{S3} for
nonzero $\beta$, and they alleviate in particular the proof of
Theorem 1. In a more refined formulation~\cite{S4}, it
asserts that the sequence

$$
0 \rightarrow S^{\prime \beta}_{\alpha}( K_1\cap
K_2) \rightarrow S^{\prime \beta}_{\alpha}(K_1)
\oplus S^{\prime \beta}_{\alpha}(K_2) \rightarrow
S^{\prime \beta}_{\alpha}(K_1\cup K_2) \rightarrow 0
\eqno{(8)}$$

\noindent
is exact. All the arrows in (8) are natural mappings and the
next to last one maps a pair of linear forms into the
difference of their restrictions. Since the involved spaces are
FS, this assertion is equivalent to saying that the dual
sequence

$$
0 \leftarrow S^{\beta}_{\alpha}( K_1\cap
K_2) \leftarrow S^{\beta}_{\alpha}(K_1)
\oplus S^{\beta}_{\alpha}(K_2) \leftarrow
 S^{\beta}_{\alpha}(K_1\cup K_2) \leftarrow 0 \eqno{(9)} $$

\noindent
is exact. Moreover both of them are topologically exact by
the same reason.

 {\it Remark 2}. The formula (8) represents a weakened version
 of a similar formula for the Fourier hyperfunctions
 ($\alpha=\beta=1$) which is valid for every pair of closed sets
 in the radially compactified ${\bf R}^n$, ensures the existence
 of supports for the elements of $S^{\prime 1}_{\,1}$ and  is
really a simple way of describing their local properties with an
 accuracy sufficient for use in QFT.

    The only nontrivial conclusion concerning  (9) is the
 exactness at $S^{\beta}_{\alpha}( K_1\cap K_2)$ which means that
 each element of this space can be
decomposed into a sum of two functions belonging to
$S^{\beta}_{\alpha}(K_j),\quad j=1,2 $. For $0<\beta<1$, such a
decomposition presents no serious problems and copies
essentially the usual partition of unity.  Namely, let $\varphi
\in H^{\beta,b}_{\alpha,a} (U) $, where $U$ is a cone-shaped
neighborhood of $K\setminus\{0\}$ with $K$ denoting the
intersection $K_1\cap K_2$.  We recall that a cone $V$ is said
to be a (relatively) compact subcone of $U$ if
$\overline{V}\setminus\{0\} \subset U$, where the bar denotes
closure, and then the notation $V\ssubset U$ is used.  Choose an
open cone $V$ so that $K\ssubset V\ssubset U$. Since the angular
distance between the closed cones $K_j \setminus V$ is nonzero,
there are open cones $V_j$ such that $K_j \setminus V \ssubset
V_j$ and

$$
|x-\xi|\geq \theta |x|,\ \quad  |x-\xi|\geq \theta |\xi|\quad
\mbox{for all} \quad x \in V_1,\> \xi \in V_2,
\eqno{(10)}
$$

\noindent
where $\theta$ is a positive constant. Let us take $\chi_0 \in
H^{\beta,b_0}_{1-\beta,a_0} $  so that $\int \chi_0 {\rm d}x
=1$ and set

$$
\chi(z) = \int \limits_{V_2}^{} \chi_0 (z-\xi){\rm d}\xi.
$$

\noindent
Using (10), one can verify that $\chi \varphi \in S^\beta_\alpha
(K_1)$ provided $a_0 < \theta/b$ and show that  $(1-\chi )\varphi
\in S^\beta_\alpha (K_2)$ if $a_0 < \theta^\prime /b$, where
$\theta^\prime$ is the angular distance of $K_2 \setminus
V $ to the complement of $V_2$. When $\beta =0$, this argument
fails because the space $S^0_1$ is trivial, but one may follow
the regular way~\cite{H1} of solving the Cousin
problem and start from a decomposition into nonanalytic
functions, using this time a standard bump function $\chi_0
\in C^\infty_0 ({\bf R}^n)$ and setting  $\chi_0 (z) = \chi_0
({\rm Re}z)$. The functions $\varphi_1 = \chi \varphi$ and
$\varphi_2 = (1-\chi)\varphi$, with $\chi$ defined as above,
possess the required behavior at infinity and, on writing
$\varphi =(\varphi_1 - \psi) + (\varphi_2 + \psi)$, our
problem amounts to finding a solution of the system of equations

$$
\frac{\partial \psi}{\partial \bar{z}_j} = \eta_j  \qquad
(j=1,\ldots, n) \eqno{(11)} $$

\noindent
with the same growth properties as those of $\eta_j = \varphi
\partial \chi /\partial \bar{z}_j$.  The existence of such a
solution is ensured by fundamental H\"{o}rmander's
theorem~\cite{H1}, though there is a subtlety here. The
point is that the function (2) is not \plu, while this
property is crucial for H\"{o}rmander's estimate. However, one
can replace $\rho$ by its greatest \plu minorant and this leaves
the space unaltered since $\ln |\varphi|$ is \plu for any
analytic function $\varphi$.

\section{Tensor Products}

We will consider the tensor product $S^0_\alpha
(U_1)\otimes_i S^0_\alpha
(U_2)$ equipped with the inductive topology $\tau_i$, that
 is, the finest locally convex topology under which the
 canonical bilinear mapping $(\varphi,\psi) \rightarrow \varphi
\otimes \psi$ is separately continuous. Recall~\cite{S} that
this topologization has the following category meaning. If
$E,F$, and $G$ are locally convex spaces and $u:\ E{\bf \times}
F \rightarrow G $ is a bilinear separately continuous mapping,
then its associated linear mapping $u_*:\ E\otimes_i F
\rightarrow G $ is continuous. Specifically, this implies the
continuity of the natural injection

$$
S^0_\alpha (U_1)\otimes_i S^0_\alpha (U_2) \rightarrow
S^0_\alpha (U_1\times U_2)
\eqno{(12)}
$$
\noindent
generated by the correspondence $\sum \varphi_j\otimes \psi_j
\rightarrow \sum \varphi_j (z_1) \psi_j (z_2)$.

{\bf Theorem 3.} {\it When extended by continuity to the
completion of the tensor product, the embedding {\rm (12)} turns
into an algebraic and topological isomorphism, that is, for any
open cones $U_1, U_2 \in {\bf R}^n$, the following
identification holds:}

$$
S^0_\alpha (U_1)\mathbin{\hat{\otimes}_i} S^0_\alpha (U_2)
= S^0_\alpha (U_1\times U_2).
$$

Our basic representation $S^0_\alpha (U) = {\rm
inj\,lim}H^{0,b}_{\alpha,a}(U) \quad  (a,b\rightarrow
\infty)$ reduces the problem to that in Hilbert spaces, but
here care is necessary because usually the topology on tensor
product of these latter is determined quite differently, by
means of its natural scalar product~\cite{RS}. It is
customary to denote the completed tensor product in the Hilbert
space category by $H_1\otimes H_2$, and we
hope this will not lead to a misunderstanding though the same
notation is used for the algebraic tensor product. The
definition (2) should be fitted to the problem at hand, and this
time we set

$$
\rho_{{}_{U,a,b}}(z)\>=\> -\sum \vert x_j/a\vert^{1/\alpha} +
b\inf_{\xi \in U}\sum \vert x_j -\xi_j \vert +
b\sum \vert y_j\vert,
\eqno{(13)}
$$

\noindent
so that the multiplicativity relation

$$
\exp\{-\rho_{{}_{U_1\times U_2}}(z_1,z_2)\} =
\exp\{-\rho_{{}_{U_1}}(z_1)\}\exp\{-\rho_{{}_{U_2}}(z_2)\}
$$
\noindent
is fulfilled.

{\bf Lemma 1.} {\it If the defining weight function is
chosen in the multiplicative form, then}

$$
H^{0,b}_{\alpha,a}  (U_1)\otimes H^{0,b}_{\alpha,a} (U_2) =
 H^{0,b}_{\alpha,a} (U_1\times U_2)
\eqno{(14)}
$$

\noindent
The proof faithfully copies that of the analogous statement
in~\cite{RS}, Sec. II.4, about weighted spaces of
locally square-integrable functions. Choosing basises
$\{\varphi_j\}$ and $\{\psi_k\}$ in the spaces on the left-hand
side of (14) and using Fubini's theorem, one makes sure that
$\{\varphi_j (z_1)\psi_k (z_2)\}$ is a basis for the space on the
right and so the natural injection of the algebraic tensor
product into it can uniquely be extended to a unitary
operator.  As an immediate consequence, one obtains that the
image of (12) is everywhere dense, by the very definition of
convergence in \saou. Thus to prove Theorem 3, it is sufficient
to show a continuous mapping

$$
S^0_\alpha (U_1\times U_2)\rightarrow S^0_\alpha
(U_1)\mathbin{\hat{\otimes}_i} S^0_\alpha (U_2)
\eqno{(15)}
$$
\noindent
whose composition with (12) is the identity mapping. Indeed,
then $\tau_i$ is the same as the topology induced on the tensor
product by that of $S^0_\alpha (U_1\times U_2)$. At this point
we will take advantage of the fact that, in the case of nuclear
Fr\'echet spaces, $\tau_i$ coincides with another often-used
topology on tensor products, the so-called topology of
equicontinuous convergence  $\tau_e$ which has a quite
simple description. The auxiliary spaces (3) are just of this
type and the space (14) can be canonically mapped into the
 completion of their tensor product endowed with the
topology $\tau_e$ by virtue of the following

{\bf Lemma 2.} {\it Let $H_1 ,H_2$ be Hilbert spaces. The
Hilbert norm $\|\cdot\|$ on their tensor product is stronger than
the norm which determines the topology $\tau_e$, that is, the
identity mapping

$$
H_1\mathbin{{\otimes}_{{}_{\|\cdot\|}}} H_2 \rightarrow
H_1\mathbin {{\otimes}_e} H_2.
\eqno{(16)}
$$

\noindent
is continuous.}

\noindent
{\it Proof.} According to~\cite{S}, for any pair of
normed spaces, the norm $\Vert \cdot\Vert_e$ is defined by

$$
\Vert \sum \varphi_j \otimes \psi_j \Vert_e =
\sup_{\|f\|^\prime \leq 1,\  \|g\|^\prime \leq 1} \sum (f,
\varphi_j)(g,\psi_j),
$$

\noindent
where $f,g$ belong to the dual spaces and the dual
norms are marked by primes. By Riesz's
theorem, in the Hilbert case the linear forms $f,g$ are
identified with elements of the spaces $H_1, H_2$ themselves and
the primes can be dropped, so the sum on the right-hand side
turns into the scalar product $(f\otimes g,\  \sum \varphi_j
\otimes \psi_j)$ and Schwarz's inequality yields $\Vert \sum
\varphi_j \otimes \psi_j \Vert_e \leq \Vert \sum \varphi_j
\otimes \psi_j \Vert$.

In order to complete the proof of Theorem 3, it remains to
combine Lemmas and note that both  topologies $\tau_e$ and
$\tau_i$ are consistent with tensoring
morphisms.  Namely, if $h_1:\  E_1 \rightarrow F_1$ and $h_2:\
E_2 \rightarrow F_2$ are continuous linear mappings of locally
convex spaces, than $h_1 \otimes h_2:\  E_1 \otimes_e E_2
\rightarrow F_1 \otimes_e F_2$ and $E_1 \otimes_i E_2
\rightarrow F_1 \otimes_i F_2$ are continuous too. For $\tau_e$,
this fact can readily be established using the explicit
form~\cite{S} of a base of neighborhoods and for $\tau_i$ by the
category arguments. First we take $h_j$ to be the natural
injections $H^{0,b}_{\alpha,a} (U_j) \rightarrow
S^{0,b+}_{\alpha,a+} (U_j) $ and use $\tau_e$. Next we consider
the inclusion mappings $S^{0,b+}_{\alpha,a+} (U_j) \rightarrow
S^0_\alpha(U_j)$ and endow the tensor products with $\tau_i$.
Making up a composition with (16) and passing to the
completions, we arrive at the embeddings

$$
H^{0,b}_{\alpha,a} (U_1\times U_2)\rightarrow S^0_\alpha
(U_1)\mathbin{\hat{\otimes}_i} S^0_\alpha (U_2)
$$
\noindent
which are evidently compatible for all $a, b$ and, taken
together, determine the desired  mapping (15)
which is continuous by the definition of inductive limit
topology, and whose restriction to $S^0_\alpha(U_1)\otimes
S^0_\alpha (U_2)$ is the identity mapping by construction.

\section{A Generalization of the Paley--Wiener--Schwartz
Theorem}
The PWS theorem whose modern presentation is available
in~\cite{H2,RS} establishes necessary and
sufficient conditions for an analytic functions to be the Laplace
transform of a tempered distribution with compact or
cone-shaped support. An analogous theorem for the more singular
distributions defined on \sab, $\beta >1$, has been derived
in~\cite{S1} with the aim of application to  strictly
 local QFT's of the Jaffe type~\cite{J1,J2}.  An appropriate
 generalization to the nonlocalizable case $0<\beta <1$ was
formulated there too, but its complete proof has been set forth
much later, see~\cite{S3}, Theorem 5.23.  In contrast to the
case of Fourier hyperfunctions
considered by Kawai~\cite{K1}, this proof is elementary in
essence and makes use of the fact that an element $f$ of
$S^{\prime \beta}_\alpha, \quad 0<\beta <1,$ is carried by a
cone if and only if the convolution $f\ast \varphi$ with
$\varphi \in S^\beta_{1-\beta}$ falls off like $\varphi$ in the
complementary cone. Now we are in a position to treat the
most difficult case $\beta =0$ of arbitrary singularity.  It is
not so easy because $S^0_1$ is trivial, but one may use a result
of Komatsu [21] who has established the growth conditions under
which analytic functions have boundary values belonging to
$S^{\prime \alpha}_0$.  Actually he considered even a finer
scale of spaces designated as ${\cal D}^{\{M_p\}}(\Omega)$,
where $M_p = p^{\alpha p}$ and $\Omega ={\bf R}^n$ for our case.
A combination of Komatsu's result with Theorems 1 and 3 leads
directly to the desired extension of the PWS theorem. We now
recall terminology and some simple facts concerning cones with
vertex at the origin. For given cone $V\subset {\bf R}^n$, the
dual cone is defined by $V^* = \{\eta:\ \eta x \geq 0,\, \forall\
x\in V\}$. The convex hull of $V$ is denoted by ch$V$.  Clearly
$V^* = ({\rm ch}V)^*$ and $V^{**} = \overline{{\rm ch}V}$. A
cone $V$ is said to be proper if $\overline{{\rm ch}V}$ does not
contain a straight line, or equivalently, if the interior of the
dual cone is nonempty. By a tubular cone in \CA is meant one of
the form ${\bf R}^n + iV$, with $V$ an open cone in ${\bf R}^n.$
We will name it simply tube and denote by $T^V$.  It should be
also noted that we define the Fourier transformation ${\cal
F}_{\xi \rightarrow x}$ of test functions by $\psi
(\xi)\rightarrow \int e^{ix\xi} \psi (\xi){\rm d}\xi$ and denote
the inverse operator by ${\cal F}_{\xi \leftarrow x}$. Their
dual operators acting on distributions are designated as ${\cal
F}_{x \rightarrow \xi}$ and $ {\cal F}_ {x \leftarrow \xi}$
respectively.

{\bf Theorem 4.} {\it Let $K$ be a closed proper cone in
\RA and let $V$ be the interior} int$K^*$ {\it of the dual cone.
Suppose that $K$ is a carrier of $f\in S^{\prime 0}_\alpha ({\bf
R}^n), \quad \alpha >1$. Then the distribution $f$ has a
uniquely determined Laplace transform $g(\zeta)$ which is
analytic in the tube $T^V $ and satisfies the estimate

$$
|g(\zeta)|\,\leq\,C_{\varepsilon,
R}(V^\prime)\, \exp\{\varepsilon\, |\eta
|^{-1/(\alpha -1)}\}\hspace{5mm} (\eta={\rm Im}\:\zeta\in
V^\prime,\ |\zeta|\leq R) \eqno{(17)} $$

\noindent
for each $\varepsilon, R >0$ and any open relatively
compact subcone $V^\prime$ of $V$.  As $\eta\rightarrow 0$
inside a fixed  $V^\prime$, the function
$g(\xi+i\eta)$ tends in the topology of $S^{\prime \alpha}_0$ to
the Fourier transform ${\cal F}_{x \rightarrow \xi}f$.

Conversely, if $g(\zeta)$ is an analytic function on
$T^V$, with $V$ an open connected cone, and if its
growth near the real boundary is bounded by {\rm (17)}, then
$g(\zeta)$ is the Laplace transform of a distribution defined on
$S^0_\alpha$ and carried by the cone $V^*$.}

{\it Remark 3}. In the case of nonzero $\beta$
examined in~\cite{S3}, the bound (17) is replaced by

$$
|g(\zeta)|\,\leq\,C_{\varepsilon
}(V^\prime)\, \exp\{ \varepsilon |\zeta|^{1/\beta}+
\varepsilon |\eta |^{-1/(\alpha -1)}\}\hspace{5mm} (\eta\in
V^\prime). \eqno{(17)^\prime} $$

 The proof of the first part of Theorem 4 is
direct and no different in the main from that for
$\beta >0$. The exponentials $e^{iz\zeta}$ belong to
\saok for each $\zeta\in T^V$ and one can set

$$
g(\zeta)\/=\/(\hat{f},\/e^{iz\zeta}),
\eqno{(18)}
$$

\noindent
with the caret denoting an extension to this space. The estimate
(17) is a consequence of the inequality

$$
|g(\zeta)|\,\leq\,\|\hat{f}\|_{U,a,b}\,\|e^{iz\zeta}\|_{U,a,b},
\eqno{(19)}
$$

\noindent
where $a, b$ can be taken arbitrarily large and $U$ is
any open cone such that $K\ssubset U$. From the definition of
the norm $\|\cdot\|_{U,a,b}$, it follows that

$$
\|e^{iz\zeta}\|_{U,a,b}\,\leq\,C_{a^\prime,b^\prime}\sup_{x,y}
\,\exp\{-x\eta-y\xi+|x/a^\prime|^{1/\alpha}-d(bx,U)-|b^\prime
y|\} \eqno{(20)} $$

\noindent
for any $a^\prime <a,\quad b^\prime <b$. Taking $b^\prime > R$,
one sees that the terms dependent on $y$ are unessential for
$|\zeta|\leq R$ and may be omitted from the formula.  The cone
$U$ and another auxiliary open cone $U^\prime$ should be taken
so that $K\ssubset U\ssubset U^\prime\ssubset
\hbox{int}V^{\prime *}$.  This is possible since $V^\prime
\ssubset \hbox{int}K^*$ implies $K\ssubset \hbox{int}V^{\prime
*}$. If $x\not\in U^\prime$, then $d(x,U)>\delta|x|$ with
$\delta>0$ and, choosing $b>R/\delta$, one can majorize the
exponent (20) by a constant.  Now let $x$ be inside $U^\prime$.
By standard compactness arguments, from the inclusion
$U^\prime\ssubset \hbox{int}V^{\prime *}$ it follows that

$$
-x\eta \leq -\theta|x||\eta| \quad \hbox{for all}\quad x\in
U^\prime, \eta\in V^\prime , \eqno{(21)} $$

 \noindent
 with   $\theta$ a positive constant. Inserting (21) into
 (20), dropping the term $d(\cdot,U)$ and locating the maximum,
we  arrive at (17).  A simple estimation proceeding along
similar lines shows that, for any $\zeta \in T^{V^\prime}$, the
difference quotients corresponding to the partial derivatives
$\partial e^{iz\zeta}/\partial \zeta_j$ converge in the topology
 of \saou and hence $g(\zeta)$ is analytic.  Further, using the
 mean value theorem, one can verify that, for each $\psi \in
 S^\alpha_0$ and $\eta \in V^\prime$, the Riemann sums
 corresponding to the integral $\int e^{iz\zeta} \psi (\xi){\rm
 d}\xi$ converge in \saou to $\varphi(z)e^{-z\eta}$, where
 $\varphi (z) = \int e^{iz\xi} \psi (\xi){\rm d}\xi$ , and
therefore the identity

$$
\int g(\zeta)\psi(\xi){\rm d}\xi = (\hat{f},
\varphi(z)e^{-z\eta})
\eqno{(22)}
$$

\noindent
holds in $V$. Finally, it is again
straightforward to prove the convergence $\varphi
e^{-z\eta}\rightarrow \varphi$ in the same space as $\eta
\rightarrow 0$ inside $V^\prime$. Thus the Fourier transform
${\cal F}_{x \rightarrow \xi}f$ is the boundary value of
$g(\zeta)$ in the sense of weak convergence. But it implies
strong convergence since $S^{\prime \alpha}_0$ is Montel. It
should be noted that the extension used in (18) is unique by
Theorem 5 proved below, but really there is no need to
appeal to it here since the difference of two
analytic functions with the same boundary value must vanish.

{\it Remark 4}. Evidently  the space of analytic functions on
$T^V$ with the growth property (17) is an algebra under
multiplication for which we will use the notation $A^\alpha
(T^V)$. It can be made into an FS space and into a
topological algebra by giving it the projective limit topology
determined by the set of norms

$$
\|g\|_{V^\prime, R,\varepsilon} = \sup_{\eta\in V^\prime,\
|\zeta|\leq R} |g(\zeta)|\exp\{-\varepsilon
|\eta|^{-1/(\alpha-1)}\}.
$$

\noindent
The above examination shows that the Laplace transformation
is a one-one and continuous mapping of  $S^{\prime 0}_\alpha (K)$
into $A^\alpha (T^V)$, $V =\hbox{int}K^*$.

We now turn to the converse assertion of Theorem 4.
Komatsu's Theorem 11.5 of~\cite{K2} ensures that every
function $g\in A^\alpha(T^V)$ has a boundary value $\hbox{b}_V
g$ which belongs to $S^{\prime \alpha}_0$ and coincides with
that in the sense of hyperfunction. In~\cite{K2}, the cone $V$
was assumed to be convex but really only its connectedness is an
essential assumption since then each $V^\prime\ssubset V$ can be
 covered by a finite enchained family of convex subcones. Our
 task is to make sure that $f =  {\cal F}_ {x
 \leftarrow \xi}(\hbox{b}_V g)$ is carried by $V^*$. We
 start with the simplest case $n=1$ and $V= {\bf R}$, when the
 dual cone $V^*$ degenerates into $\{0\}$.  Then $\hbox{b}_V g$
 is merely the restriction of the entire function $g$ to the
 real axis and $f$ takes the form $\sum c_k
\delta^{(k)}(0)$, where $\overline{\lim}|c_k|^{1/k}= 0.$ Let
 $\varphi \in S^0_\alpha(\{0\})$ and so $\Vert\varphi\Vert^2_b =
\int |\varphi|^2 \exp(-2b|z|)\,{\rm
d}\lambda <\infty$ for some $b$. Denote
by $\chi_{{}_R}$ the characteristic function of the disk $|z|<R$
smoothed by convolution with a $C^\infty$ function whose support
is contained in the unit disk, and apply the
Cauchy--Green integral formula to $\varphi\chi_{{}_R}$. Using
Schwarz's inequality and taking into account the support
properties of $\partial\chi_{{}_R}/\partial\bar{z}$, we find that

$$
 |\varphi^{(k)}|\leq C\,\Vert\varphi\Vert_b\, k!\, (R-1)^{-(k+1)}
e^{b(R+1)} \sqrt{R}.
$$

 \noindent
 The optimization of $R$, combined with Stirling's
 formula, yields the estimate $|\varphi^{(k)}|\leq
 C\,\Vert\varphi\Vert_b\, b^k$ which shows that the above series
 does determine a continuous functional on $S^0_\alpha(\{0\})$.
Next we consider the case $V= {\bf R}_+,\; V^* = \overline{{\bf
R}}_+ $ which is slightly more involved.  This time we make use
of Theorem 1 and decompose $f$ into a sum of two distributions
$f_+,\  f_-$ carried by $\overline{{\bf R}}_+$ and
 $\overline{{\bf R}}_-$. On doing the Laplace transformation, we
get $g(x+i0) =g_+(x+i0) + g_-(x-i0)$ with $g_\pm \in
A^\alpha({\bf C}_\pm)$.  By the "edge of the wedge"
theorem~\cite{H2}, there is an entire function which continues
both $g-g_+$ and $g_-$ and hence $f-f_+$ is carried by the
origin according to what has just been said.  This completes the
proof for $n=1$.  When $n\geq 2$, we may assume without loss of
generality that the first  basis vector lies in
$V$.  Let us take $\psi \in S^\alpha_0 ({\bf R}^{n-1})$,
introduce the designation $\check{\xi} = (\xi_2,\ldots, \xi_n)$,
and consider the mapping

  $$
  \psi  \rightarrow g_1 (\zeta_1) =
  \int g(\zeta_1, \check{\xi}) \psi
  (\check{\xi})\,d\check{\xi}
  $$

  \noindent
  of  $S^\alpha_0 ({\bf R}^{n-1})$ into $A^\alpha ({\bf
  C}_+)$. As can easily be seen, it is continuous. Let $f_1$
   be the distribution in $S^{\prime 0}_\alpha (\overline{{\bf
R}}_+)$ whose Laplace transform is $g_1$. Note that
    $S^{\prime 0}_\alpha (\overline{{\bf R}}_+)$ is no different
   from $S^{\prime 0}_\alpha ({\bf R}_+)$.  The correspondence
   $g_1\rightarrow f_1$
  is also continuous by virtue the open mapping theorem in
   Grothendieck's version which covers the FS spaces. Since the
   Fourier transformation determines an isomorphism of
  $S^\alpha_0({\bf R}^{n-1})$ onto $S_\alpha^0({\bf R}^{n-1})$ ,
  we thereby obtain a bilinear separately continuous form on
  $S^0_\alpha({\bf R}_+)\times S_\alpha^0({\bf R}^{n-1})$  and
  hence, by Theorem 3, a linear continuous form on
   $S^0_\alpha(H_1)$, where $H_1$ is the half-space $x_1 >0$.
   Its restriction to $S^0_\alpha({\bf R})\otimes
   S_\alpha^0({\bf R}^{n-1})$  coincides obviously with that of
   $f$ and, since this tensor product is dense in
   $S_\alpha^0({\bf R}^n)$ by Theorem 3 again, we infer that
   $\overline{H}_1$ is a carrier of $f$. The same argument shows
   that $f$ is carried by every half-space $\{x:  x\eta\geq 0\}$
   with $\eta \in V$, whose intersection is just the cone $V^*$.
   To complete the proof of Theorem 4, it remains to apply the
   last conclusion of Theorem 1.

   {\bf Corollary 1.} {\it Every function analytic in a tubular
   connected cone $T^V$ and satisfying the condition {\rm (17)}
   allows an analytic continuation into the tube $T^{{\rm ch}V}$
   which possesses the same growth property.}

   {\bf Corollary 2.} {\it If a closed cone K is convex and
   proper, then the distributions defined on \sao and carried by
   $K$ form a topological algebra under convolution, and the
   Laplace transformation maps it isomorphically onto the
   topological algebra $A^\alpha(T^V),\quad V={\rm int}K^*$.}

   {\it Proof}. For brevity, we will identify the set of
   distributions carried by $K$ with the space $S^{\prime
   0}_\alpha(K)$. This is admissible due to Theorem 5
 proved below. As shown in Theorem 5.16 of~\cite{S3}, for each
   $f_0\in S^{\prime 0}_\alpha(K)$ and $\varphi\in
   S^0_\alpha$, the convolution $(f_0\ast
   \varphi)(x) = (f_0, \varphi(x-\cdot))$  belongs to \saou, where $U$ is any open cone
   compact in the complement ${\sf C}K$ of $K$. (In the context of
   axiomatic QFT, a similar assertion for tempered
   distributions is sometimes referred to as Hepp's lemma.)
   Moreover, the mapping \sao$\rightarrow$\saou:$\
   \varphi\rightarrow f_0\ast \varphi$ is continuous. It should
   be noted that this fact is true for arbitrary close cone $K$
   and is proved in~\cite{S3}  even for more general classes of
   distributions. It enables one to define the
   convolution $f\ast f_0$ for each $f$ carried by a compact
   subcone $C$ of $-{\sf C}K$.
     Namely, the mapping
   $S^{\prime 0}_\alpha(C)\rightarrow S^{\prime 0}_\alpha :\
   f\rightarrow f\ast f_0$ is defined to be dual of
   \sao$\rightarrow S^0_\alpha(C):\ \varphi\rightarrow (f_0,
   \varphi(x+ \cdot))$.  If $K$ is proper, then it is itself
   compact in $-{\sf C}K$, and we obtain the bilinear mapping
   $S^{\prime 0}_\alpha(K)\times S^{\prime
   0}_\alpha(K)\rightarrow S^{\prime 0}_\alpha$.
    By the estimates obtained
   in~\cite{S3}, it is continuous in the second argument as well
   as in the first one and, since the space  $S^{\prime
   0}_\alpha(K)$ is Fr\'echet, this separate continuity implies
   continuity, see~\cite{S}, Theorem III.5.1. When being
   restricted to the Schwartz distributions of compact support,
   the above convolution operation is in line with the
   definition of~\cite{GS} and hence corresponds to the
   multiplication of the Laplace transforms by the usual PWS
   theorem. Taking into account analyticity of the elements of
   $S_\alpha^0 (K)$ and reflexivity of this space and
   using the Hahn--Banach theorem, one sees
   immediately that these distributions and even those
   supported by the origin are dense in $S^{\prime 0}_\alpha(K)$.
   Therefore, by virtue of Theorem 4 and the open mapping
   theorem, this correspondence holds true for the
   whole of $S^{\prime 0}_\alpha(K)$ and in particular
   the convolution does not take distributions out
   this space. This completes the proof. It is
   worthwhile to note that, for the tempered
 distributions with support contained in a proper closed convex
 cone, the convolution is defined in~\cite{H2} through another
   construction which demonstrates  associativity
   and commutativity of this operation but involves a
   localization and so is inapplicable to our more
   general case. However the PWS theorem enables these
   algebraic properties to be interpreted as those of
   multiplication.

   \section{Approximation Theorem}

   In order to develop the calculus of interest to us by analogy
   with that of hyperfunctions, we need a suitable approximation
   theorem for functions belonging to \saok.
   However, unlike the case~\cite{S3} of nonzero $\beta$,
   the customary means of approximation such as smoothing and
   cutoff are insufficient here, except for the degenerate cone
   $K=\{0\}$. Because of this, we will follow the line of
   thought used in~\cite{H1}. Before showing the desired
   theorem for \saok, we consider a more general situation when
   the defining function $\rho (z)$ need not be of the
   form (2), and we denote by $H_\rho$ the closed subspace of
   $L_\rho^2 = L^2( {\bf C}^n, e^{-\rho } {\rm d}\lambda)$
   consisting of analytic functions. If $\rho$ is \plu, then
   $\hat{\rho}$ denotes the strictly \plu function $\rho+ 2\ln
   (1+|z|^2)$. It should be noted that the proof presented below
   is based on exploiting $L^2$--estimates for solutions of
   the dual equation rather than those for the equations (11)
   themselves, in contrast to the strategy sketched
   in~\cite{S4}. We begin with quoting H\"{o}rmander's result
   derived in~\cite{H1}, Sec.4.4.  It was not stated there in
     the form of a theorem but this may be done as follows.

   Let $\rho$ be a  \plu $C^2$ function on \CA and let $v\in
   L^2_{\hat{\rho}}$. If $v$ is orthogonal to each analytic
   function in this space, then the equation

   $$
   \sum \frac{\partial (h_je^{-\hat{\rho}})}{\partial z_j} = -v e^{-\hat{\rho}}
   \eqno{(23)}
   $$
   \noindent
   has a solution satisfying the estimate

   $$
   2\int |h|^2(1+|z|^2)^{-2}e^{-\hat{\rho}}{\rm d}\lambda \leq
   \int |v|^2 e^{-\hat{\rho}} {\rm d}v.
   \eqno{(24)}
   $$
   \noindent
   This fact enables one to prove the following lemma.

   {\bf Lemma 3}. {\it Let $\rho_0,\:\rho$ and $\rho^\prime$
be continuous real valued functions on \CA such that
    $\rho_0\leq \rho^\prime,\:\rho \leq
   \rho^\prime$ and hence $H_{\rho_0},\: H_\rho
   $ can be regarded as vector subspaces of
   $H_{\rho^\prime}$. Suppose that there exists a sequence
   of smooth \plu functions $\rho_\nu,\quad \nu=1,2,...$, which
   satisfy the conditions:

   \begin{description}
   \item[(i)] $\, \ \, \hat{\rho}_\nu \leq \rho^\prime,$
   \item[(ii)]$\, \hat{\rho}_\nu \leq \rho_0 + C_\nu,$
   \item[(iii)]$\rho_\nu \geq \rho \quad$ for $\;|z|\leq \nu$.
   \end{description}
   Then $H_{\rho_0}$ is dense in the space $H_\rho$ under the
   topology of $H_{\rho^\prime}$.

   Proof}. The closure of $H_{\rho_0}$ in $H_{\rho^\prime}$
   covers $H_\rho$ if and only if $H_{\rho_0}^\bot \subset
   H_\rho^\bot$, where the orthogonal complement is determined
   by the scalar product of $H_{\rho^\prime}$. It is this
   inclusion that we shall prove. Let $v\in H_{\rho^\prime}$ and
   let $\int \bar{v} u \exp\{-\rho^\prime\}{\rm d}\lambda = 0$
   for each $u\in H_{\rho_0}$. It suffices to derive the
   representation

   $$
   -ve^{-\rho^\prime}= \sum \frac{\partial w_j}{\partial
   z_j},
   \eqno{(25)}
   $$
   \noindent
   with $w_j\in L^2_{-\rho}$, which should be fulfilled in the
   sense of Schwartz distribution, that is,

   $$
   \int \bar{v} \varphi e^{-\rho^\prime}{\rm d}\lambda = \sum
   \int \bar{w}_j\frac{\partial \varphi}{\partial \bar{z}_j}{\rm d}\lambda
   \quad \mbox{for all }\varphi \in C^\infty_0.
   \eqno{(26)}
   $$
   \noindent
   In fact, let $u$ be an element of $L^2_\rho$ such that
   $\bar{\partial}_ju \in L^2_\rho$ for every $j$. In the
   ordinary way which combines a cutoff with smoothing and is
   used, e.g., in the proof of Lemma 4.1.3 of~\cite{H1}, one
   can approximate $u$ by $C^\infty_0$ functions
   in the norm
   $\|\cdot\|_\rho + \|\bar{\partial}(\cdot)\|_\rho$.
   If  $u\in H_\rho$, then by passing to the limit in (26),
   one obtains immediately $\int \bar{v} u
   \exp\{-\rho^\prime \}{\rm d}\lambda = 0$.

       Let us now rewrite the orthogonality condition $v\bot
   H_{\rho_0}$ as follows

   $$
    \int \bar{v}
    e^{-\rho^\prime +\hat{\rho}_\nu}ue^{-\hat{\rho}_\nu}{\rm
    d}\lambda = 0.
   $$
   \noindent
   By virtue of (i), the function
   $v\exp\{-\rho^\prime +\hat{\rho}_\nu\}$ belongs to
   $L^2_{\hat{\rho}_\nu}$. Furthermore, it is orthogonal to
   each  analytic element of
   $L^2_{\hat{\rho}_\nu}$ since these are contained in
   $H_{\rho_0}$ due to (ii). Thus by H\"{o}rmander's theorem
   referred to, the equation

    $$ \sum \frac{\partial
   (h_je^{-\hat{\rho}_\nu})}{\partial z_j} = -v e^{-\rho^\prime}
   $$

   \noindent
   has a solution such that

   $$
   2\int |h|^2(1+|z|^2)^{-2}e^{-\hat{\rho}_\nu}{\rm d}\lambda
   \leq \int |v|^2 e^{-2\rho^\prime +\hat{\rho}_\nu} {\rm
   d}\lambda.
   $$

   \noindent
   Setting $h_j\exp\{-\hat{\rho}_\nu\}=w^\nu_j$ and using (i) again, we
   get a family of representations of the form (25), where

   $$
   2\int |w^\nu|^2 e^{\rho_\nu}{\rm d}\lambda \leq \int
   |v|^2 e^{-\rho^\prime}{\rm d}\lambda .
   \eqno{(27)}
   $$

   \noindent
      Let $w^\nu_{\rm{cut}}(z)=w^\nu (z)\chi(z/\nu)$, where $\chi
   \in C^\infty_0$ is a standard cutoff function with
   support in the unit ball and equal to 1 in a neighborhood of
   the origin.  Owing to the condition (iii), since
   the right-hand side of (27) is independent of $\nu$,
    the sequence $ w^\nu_{\rm{cut}}$ is strongly
   bounded in $L^2_{-\rho}$ and one can draw from it a weakly
   convergent subsequence. We take $w$ to be its limit. By
   construction, $w^\nu_{\rm{cut}}$ coincides with $w^\nu$ on
   every given compact set when $\nu$ is large enough and so $w$
   does satisfy the equation (26). This ends the proof.

   We shall apply Lemma 3 to the triple
$H^{0,b_0}_{\alpha,a_0},\;
H^{0,b}_{\alpha,a}(U),\;
H^{0,b^\prime}_{\alpha,a^\prime}(U)$ with $a_0=a^\prime$ and
   $b_0= b^\prime$ for simplicity. The required sequence
$\rho_\nu$ will be constructed   starting from auxiliary
functions of the form $\ln|\varphi_{{}_N}|$, where
$\varphi_{{}_N}$ belong to $S^0_\alpha$ and are bounded by

$$
|\varphi_{{}_N}| \leq A\,\exp\{|y|-|x/\gamma|^{1/\alpha}\},
\eqno{(28)}
$$

\noindent
with constants $A,\,\gamma$ common to all $\varphi_{{}_N}$.

   {\bf Lemma 4}. {\it For each $\alpha>1,\gamma>0$ and
$0<\sigma< 1/2$,  there exists a family of functions
   $\varphi_{{}_N}(z)\in S^0_\alpha({\bf R}),\quad N=1,2,...$,
which satisfy the bound {\rm (28)} and the following additional
requirements:

   \begin{description}
   \item[(*)] $\,\ \ln|\varphi_{{}_N}(z)| \geq
   \sigma|y|  \qquad \, \,$for$ \  |x|\leq 1$,
 \item[(**)] $ \ln|\varphi_{{}_N}(z)| \leq |y|-
    N\ln^+ (\sigma|x|/N) +B$,
    \end{description}
    where $\ln^+r=\max(0,\:\ln r)$ and B is a constant
   independent  of $N$.

   Proof}. Let $\chi(t)$ be the characteristic function of the
   interval [-1,\,1], and let

   $$
   \chi_{{}_N} (t) = \frac{N}{2}\int\limits^{t+1/N}_{t-1/N}{\rm
   d}t_N\ldots
   \frac{N}{2}\int\limits^{t_2+1/N}_{t_2-1/N}\chi(t_1)\,{\rm
   d}t_1.
   \eqno{(29)}
   $$

   \noindent
     It is well known that the inequality $\alpha>1$ is a
   non-quasianaliticity condition under which, for any
   $\gamma,\,\delta >0$, one can construct (by an iteration
   procedure similar to (29)) a nonnegative even function
   $\omega$ such that

   $$
   \|\omega^{(k)}(t)\|\leq A\gamma^k \,k^{\alpha k}, \quad \int
   \omega\, {\rm d}t =1,\quad \hbox{supp}\:\omega \subset
   [-\delta,\,\delta].
    \eqno{(30)} $$

   \noindent
   It is easy to see that the Laplace transform of
   $\psi_{{}_N}=\chi_{{}_N}\ast\omega$ satisfies all the stated
   requirements after a suitable rescaling. First, dropping for
the moment the subscript, we note that $\chi^{(N)}$ is of the
   form $(N/2)\sum (-1)^i \chi(t-\tau_i)$, where the sum
   involves $ 2^N$ terms, and so is dominated by $N^N$.
   Thus, making use of Kolmogorov's inequalities $M_k\leq
   2M_0^{1-k/N} M_N^{k/N}$ for modulus maxima of successive
   derivatives, we can write

   $$
   \|\chi^{(k)}(t)\|\leq 2\,N^k \quad (k\leq N),
   \quad \int \chi\,{\rm d}t =2,\quad \hbox{supp}\:\omega
   \subset [-2,\,2],  \eqno{(31)} $$

   \noindent
   with the last two properties being evident. From (30), (31),
   it follows the estimate

$$
|x^k\tilde{\psi}(z)|\leq
\int\limits_{-2-\delta}^{2+\delta}\left| e^{izt}
\psi^{(k)}(t)\right|\,{\rm d}t\leq
C_\delta \,e^{(2+\delta) |y|}\gamma^k \,k^{\alpha k}
 $$

\noindent
which implies (28) upon changing from $z$ to $z/(2+\delta)$ and
redefining $A,\,\gamma$, because $\inf_k k^{\alpha k}/r^k\leq
C\,\exp\{-(\alpha/e)|r|^{1/\alpha}\}$. After the replacement
$\gamma^k \,k^{\alpha k}\rightarrow N^k$, the same estimate and
rescaling yield (**). Note now that, for $|x| <1/2$ and $\delta$
small enough, the inequality $\cos xt > 1/2$ holds on
supp$\,\psi$ and that $\psi$ is an even function with the
properties $0\leq \psi \leq 1, \quad \int\psi {\rm d}t =2$.
Therefore

   $$
   |\tilde{\psi}(z)|\geq \int\, e^{-yt}\cos xt\,
   \psi(t)\,{\rm d}t\geq
   {1\over 2}\int
\limits^{2+\delta}_{1-\delta}e^{-yt} \psi(t){\rm d}t\geq
{\delta\over2}e^{(1-\delta)|y|}.
 $$

\noindent
Thus, if $\delta$ is sufficiently close to zero, the function
$\varphi_{{}_N} (z)=(2/ \delta){\tilde{\psi}}_{{}_N} (z/(2+\delta))$
satisfies all the conditions required.

{\it Remark 5}. Besides the approximation theorem below, we
would like to point out another application~\cite{S2} of the
multipliers $\varphi_{{}_N}$. Due to the condition (**), the
lower envelope $\inf_{{}_N} |\varphi_{{}_N}(x)$ decreases
   exponentially when $x$ approaches infinity, and this makes it
   possible to improve considerably Ruelle's original derivation
of the cluster decomposition properties in QFT. Namely, when
combined with Hepp's lemma, the above construction shows rather
directly the exponential character of this property in field
theories having a mas gap without using any more special tools,
and such a simple derivation is applicable to quantum fields of
arbitrary singularity including nonlocal ones.

In what follows, we use the specification (13)
   which again is most suitable.

{\bf Theorem 5.} {\it Let $\alpha >1$ and let $U$ be an open
cone in \RA. For any $a^\prime>a$ and $b^\prime >2enb$, the
 space $H^{0,b^\prime}_{\alpha,a^\prime}$ is dense in
$H^{0,b}_{\alpha,a}(U)$ under the topology of the space
$H^{0,b^\prime}_{\alpha,a^\prime}(U)$ which contains both of
them. As a consequence, the space \sao is
dense in $S_\alpha^0(U)$  and all the more it is dense in each
space \saok, where $K$ is a closed cone. }

{\it Proof}.  We may
assume that  $b^\prime>1$ and $b=\sigma/en$ with some
   $\sigma<1/2$ since the problem is reduced to this particular
   case by rescaling, and then apply Lemmas 3 and 4. Let us
   denote the number $\sigma/en$ by $\sigma^\prime$. It is
sufficient to find a sequence of \plu functions such that

$$
\rho_\nu \leq \rho_{{}_{U,a,b^\prime}} \quad  \hbox{and}\quad
 \rho_\nu \leq \rho_{{}_{{\bf R}^n,a,b^\prime}}+C_{\nu}\quad
\hbox{everywhere}, \eqno{(32)} $$

\noindent
 and

$$
 \rho_\nu \geq \rho_{{}_{U,a,\sigma^\prime}}-C \quad
\hbox{for}\quad \sum |x_j|<\nu, \eqno{(33)} $$

\noindent
with $C$ independent of $\nu$.
 In fact, if
 $\rho_\nu$ is not smooth, then one can correct this defect
 forming  the convolution by a nonnegative $C_0^\infty$ function
$\chi(|z|)$ such that $ \int \chi {\rm d}\lambda=1$,
which preserves plurisubharmonicity, see~\cite{H1}. Elementary
 estimates using the triangle inequalities (6) show that the
 convolution satisfies the same conditions  with some additional
 constants. Furthermore, we have $\ln(1+|z|^2)\leq
\delta|x|^{1/\alpha}+\delta|y|+C_\delta$ with arbitrarily small
 $\delta$ which can be included in $a^\prime,b^\prime$. Thus,
 (32) and (33)   ensure the fulfillment of all the conditions of
 Lemma 3 for the triple $\rho_{{}_{{\bf
 R}^n,a^\prime,b^\prime}},\;
\rho_{{}_{U,a,\sigma^\prime}}-C^\prime,\;
 \rho_{{}_{U,a^\prime,b^\prime}}+C^{\prime \prime}$ which
 defines the same spaces for any values of the constants.

  Let us denote by $\varepsilon$ the difference $b^\prime-1$
 and set $\gamma =\varepsilon a$ in (28). Let $\varphi_{{}_N}$ be
 functions whose existence in $S^0_\alpha({\bf R})$
 is established by Lemma 4 and let
  $\Phi_N=\sum\ln|\varphi_{{}_N}(z_j)|,\quad  \Phi
 =\sum \ln\varphi_1(\varepsilon z_j)|$.
 We define $\rho_\nu$ by

 $$
 \rho_\nu(z)=\sup_{|\kappa|\leq\nu}\{\Phi (z-\kappa) +
 L(\kappa)\} +\sup_{|\kappa|\leq\nu}\{\Phi_N
 (z-\kappa) + L_N(\kappa)\}, \eqno{(34)} $$

 \noindent
 where $\kappa$ runs over real multi-integers, $|\kappa|=\sum
 |\kappa_j|$ and  $L,\;L_N$  are
 the least upper bounds of those $l$ for which

 $$
 \Phi (z-\kappa) +l\leq -\sum \left|{x_j\over
 a}\right|^{1/\alpha} + \varepsilon\sum |y_j|\,\ {\rm
 and}\,\ \Phi_N (z-\kappa) +l\leq \inf_{\xi\in U}\sum
 |x_j-\xi_j| + \sum |y_j| \eqno{(35)} $$
 \noindent
 respectively. Since
 only a finite number of $\kappa$'s is involved in (34), the
 function $\rho_\nu$ is surely \plu and it satisfies
   (32) due to the bound (28) and by construction
   since the
 sum of right-hand sides of the inequalities (35) is just
 $\rho_{{}_{U,a,b^\prime}}$.
  The parameter $N$ in
 (34) is regarded as a function of $\kappa$ which should be
 chosen in such a way to satisfy (33).

Let $|x|<\nu$. Note  that $\Phi(z-\kappa)\geq 0$ for
 $|x_j-\kappa_j|\leq 1/\varepsilon$ by the condition (*) and
 $L(\kappa)\geq -\sum|\kappa_j/a|^{1/\alpha} -n\ln A$ by
 definition of $L$ and due to (28). Therefore, if  $\kappa_j$'s
  are equal to  the integer parts
of $x_j$'s, then
 $\Phi(z-\kappa)+L(\kappa)\geq -\sum|x_j/a|^{1/\alpha}-C_1$.
 With the same $\kappa_j$'s we have $\Phi_N(z-\kappa)\geq
 \sigma^\prime\sum |y_j| $ since $1/2 \geq \sigma^\prime$. Thus
 to complete the proof, we only need to show that
 $L_N(\kappa)\geq \sigma^\prime\inf_{\xi\in
 U}\sum|\kappa_j-\xi_j|-C_2$, with $C_2$ independent of
 $\kappa$, providing $N(\kappa)$ is properly chosen. Then (33)
 is fulfilled for $C=n\sigma^\prime+C_1+C_2$.  We assume $\kappa
\notin U$, for otherwise the infimum is zero and this inequality
is obviously valid with $C_2=n\ln A$,  and we now pass to the
 Euclidean norm $|x|$ through the use of

 $$
 \sum \ln^+|x_j|\geq \ln^+ {|x|\over \sqrt{n}},\quad
 |x|\leq\sum |x_j|.
  $$

  \noindent
  Then it
 remains to examine the inequality

 $$
 -N\ln^+{\sigma|x|\over N\sqrt{n}}+l\leq
 \inf_{\xi\in U}|x+\kappa-\xi|
  \eqno{(36)}
  $$

  \noindent
 which implies the second of
inequalities (35) by virtue of (**), upon replacing $l$ by
 $l+nB$ and $x$ by $x-\kappa$. Let $d(\kappa)$ be the Euclidean
 distance of the point $\kappa$ to $U$.  The infimum on the
 right-hand side of (36) can be minorized by that
taken over $\{\xi:\ |\kappa-\xi|\geq d\}$.  The latter is equal
to $d-r$ when $r\equiv |x|\leq d$ and to zero otherwise. Thus we
 face an easy task to compare a piecewise convex function with a
 piecewise linear one.  Regarding $N$ for the moment as a
 continuous parameter and carring out the minimization
  of $-N\ln |\sigma r/ N\sqrt{n}|$ with respect to $N$
 at the point $r=d$, we find  $N(\kappa) =\sigma
 d/e\sqrt{n}$.  Next we equate the left-hand side of (36) to
 zero at the same point and obtain $l=N$.  It is readily
 verified that then $l<d-r$ at the break point of $\ln^+|\sigma
 r/ N\sqrt{n}|$ and hence the inequality (36) holds
 everywhere. If $N$ and $l$ are  equal to the integer part of
 $\sigma d/e\sqrt{n}$, then it is also fulfilled
 and, returning to the norm $\sum |x_j|\leq \sqrt{n}|x|$, we
 arrive at the estimate

 $$
 L(\kappa)\geq {\sigma d(\kappa)\over
 e\sqrt{n}}-1-nB \geq \sigma^\prime\inf_{\xi\in U}\sum
 |\kappa_j-\xi_j|-C^\prime
 $$
 \noindent
 which completes the proof.

\section{Conclusion}
The purpose of this work was to establish the basic properties of
those distribution classes which provide the widest framework
for constructing quantum field models with singular infrared
behavior. The employment of such distributions is to keep
analytical tools of QFT in coordination with the indefinite
metric formalism in going beyond perturbation theory. We have
shown that a considerable part of Schwartz's theory of
distributions and Sato's theory of hyperfunctions has
interesting analogues under arbitrary singularity, and it is
luckily just this part that is of use in QFT. The results
obtained form a basis for further developments which are beyond
the scope of the present paper, such as general structure
theorems and special ones concerning Lorentz covariant
distributions, an invariant splitting of distributions carried
by the closed light cone, a representation of the
Jost--Lehmann--Dyson type, a connection with the concept of wave
front and so on. These topics are under investigation and both
Theorems 4 and 5 are of prime importance therein.

It is noteworthy that similar theorems are valid for
distributions defined on the space $S^0=S^0_\infty$. This is of
interest particularly in view of  L\"{u}cke's
works [18-19] which show a way of deriving
 the connection between spin and statistics and the {\it
TCP}--invariance for nonlocal fields whose matrix elements
belong to $S^{\prime 0}$ and so have arbitrary high-energy
behavior. This time a part of argument is even simpler since
$S^0$ is none other than the Fourier transform of Schwartz's space
{\it D}, and in proving the analogue of Theorem 4 one can appeal
 to the text~\cite{H2} instead of Komatsu's theorem. However,
 the topological structure of $S^0(K)$ is rather complicated and
 in this respect exploiting the distributions of wider class
 $S^{\prime 0}_\alpha$ is perhaps preferable here too. We would
like also to point out that Theorem 4 has an immediate
application to the problem of formulating the generalized
spectral condition for infrared singular quantum fields raised
by Moschella and Strocchi~\cite{MS} and enables one to cope with
it in a manner completely analogous to that used in nonlocal QFT
for generalization of the microcausality condition. Really, as
 argued in more detail in an accompanying paper, the first part
of the generalized Paley--Wiener--Schwartz theorem  establishes
general bounds on the correlation functions of gauge fields
while the second one specifies the test function spaces which
correspond to quantum fields with given infrared behavior.

{\it Acknowledgments}. This work was supported in part by
 the Russian Foundation for Fundamental
Research under Contract   No. 93-02-03379, and in part by
 Grant No.  INTAS-93-2058. The author is also grateful to Professor
V.~Ya.~Fainberg for helpful discussions.

\vspace{0.5cm}

\xipt

\end{document}